\documentclass[%
 aip, jcp
 amsmath,amssymb,
 reprint,%
]{revtex4-1}

\usepackage{graphicx}
\usepackage{dcolumn}
\usepackage{bm}
\usepackage{subfig}

\usepackage[utf8]{inputenc}
\usepackage[T1]{fontenc}
\usepackage{mathptmx}
\usepackage{etoolbox}

\makeatletter
\def\@email#1#2{%
 \endgroup
 \patchcmd{\titleblock@produce}
  {\frontmatter@RRAPformat}
  {\frontmatter@RRAPformat{\produce@RRAP{*#1\href{mailto:#2}{#2}}}\frontmatter@RRAPformat}
  {}{}
}%
\makeatother
\begin{document}

\preprint{AIP/123-QED}

\title[Lithium Adsorption on Polyacenes and Zig-zag-edge Graphene Nano-Strips]{Lithium Adsorption on Polyacenes and Zig-zag-edge Graphene Nano-Strips}

\author{Yenni P. Ortiz}
\email{yortiz@icf.unam.mx}
\affiliation{Centro Internacional de Ciencias, A.C., Cuernavaca, Morelos, M\'exico}

\author{Douglas J. Klein}
\affiliation{MARS, Texas A\&M University at Galveston, Texas, USA}

\author{Thomas H. Seligman}
\affiliation{Centro Internacional de Ciencias, A.C., Cuernavaca, Morelos, M\'exico}
\affiliation{Instituto de Ciencias F\'isicas, Universidad Nacional Aut\'onoma de M\'exico, Cuernavaca, M\'exico}

\date{\today}

\begin{abstract}
 The effect of increased electron-density from adsorbed Li atoms in polyacenes and in nano-ribbons with zig-zag edge is discussed from three different points of view: first in terms of resonance theoretical considerations, second in terms of H\"uckel-theoretic edge-localized frontier molecular orbitals and third using density functional theory (DFT) for anthracene, for polyacene polymers, and for graphene strips. 
Concurrence of the three approaches in some relevant aspects leads to relevant conclusions for zig-zag edge graphene.
\end{abstract}

\maketitle

\section{Introduction}
	A diverse range of intercalates \cite{intera,interb,interc} in graphite (and charcoal) have long been known, including those involving alkali metals.  With the emergence of single-sheet graphene \cite{novo}, adsorption on such sheets is of interest, say as to how adsorbed alkali metals modify the exceptional electronic structure near graphenic zig-zag edges\cite{alkgraph1, alkgraph2, alkgraph3}. It is understood that reactivity should be greatest where there are unpaired spins, as indeed occurs on such zig-zag edges.\\

In recent years, computational and experimental studies have highlighted the crucial role of lithium adsorption in tuning the electronic and mechanical properties of two-dimensional carbon materials. Modern ab initio analyses \cite{aljaber2025, gong2021} have shown that Li binding energies are substantially enhanced at zig-zag edges and defect sites, where charge transfer and local rehybridization lead to metallic or magnetic behavior. These findings confirm that the concepts first developed for polyacenes-such as resonance stabilization, edge localization, and partial sp\textsuperscript{3} distortion-remain valid descriptors for Li-carbon interactions in extended systems. Moreover, lithium adsorption on graphene and related nanostructures is now recognized as a key mechanism underlying their performance as anode materials in Li-ion batteries \cite{aljaber2025}.\\

There is special interest in Li atoms adsorption on benzenoids, say on polyacenes, where there is an anticipated 0-band gap (in the high-polymer limit \cite{highpol}), much as for graphene or more so for graphene edges. All this fits into a general understanding of functionalizing graphene as is anticipated to be crucial for prospective nanotechnologycal properties. Here, then we address alkali metals (especially lithium) adsorption first on polyacenes, semi-infinite polyacenes, then on graphene strips with the same sort of boundaries as the polyacenes and semi-infinite graphene strips. Our approach is 3-fold:
	
	\begin{itemize}
  \item via classical chemical resonance theory
   \item via simple H\"uckel molecular orbital theory (simple tight-binding) argumentation
   \item via numerics using modern DFT
   \end{itemize}
This emphasizes some predicted commonalities, so that the predictions should be more reliable and also indicate what sorts of things can be adduced from the simple qualitative theories.  Indeed, when it comes to electron pairing, including both weak pairing and the completely unpaired limit, a simple qualitative resonance-theory approach seems to work quite well.  

\section{Li Adsorption on Finite Polyacenes}

\begin{figure}[htbp]
\begin{center}
\includegraphics[width=.4\textwidth]{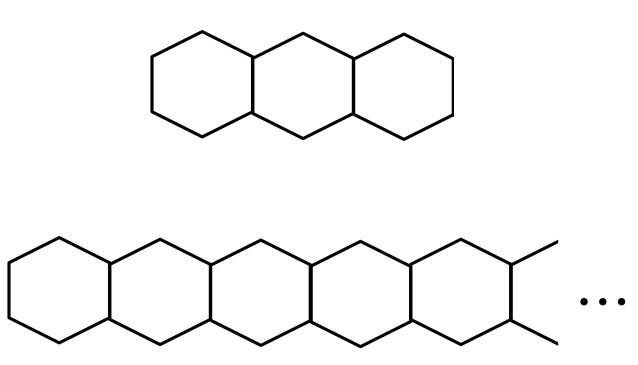}\caption{First line: The $\pi$-network for anthracene. Second line: The $\pi$-network for semi infinite polyacene}
\label{pinet}
\end{center}
\end{figure}

	We consider each of these three approaches separately, then compare the different predictions.  The first two schemes are especially simple, without computer calculations, and are hoped to be qualitatively correct to aid in understanding what goes on in a variety of situations, including Li adsorption on graphene.  For the polyacenes, we start from a standard geometry of a chain of regular hexagons, and address perturbations from this regard due to adsorbed Li atoms.  Our smallest considered polyacene is that of 3 hexagonal rings (anthracene), with a network (graph) as indicated in Fig. \ref{pinet}.  There is, one attached (but unshown) H atom at each coordination 2 (degree-2) vertex.

\subsection{Chemical Resonance-Theoretic Argument}

\begin{figure}
        \centering
                 \includegraphics[width=.4\textwidth]{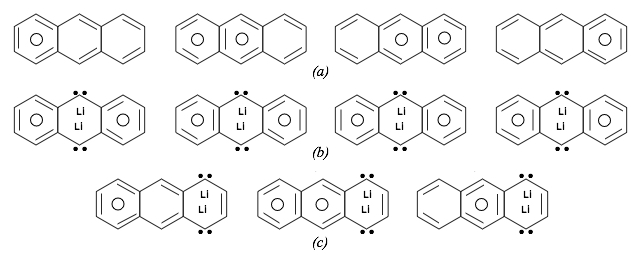}   
     \caption{ Neighbor-paired resonance structures for (neutral) anthracene in $(a)$. Resonance structures for Lithium coordinated anthracene in central ring in $(b)$ and in end ring in $(c)$. Benzene-like local conjugated 6-cycles are indicated with a small circle in the center of the associated ring.}\label{resonances}
\end{figure}

This follows Pauling's \cite{pauling} and Wheland's \cite{reson} ideas that for anthracene there are 4 (neighbor-paired) Kekul\'e structures, with an average of 1.5 conjugated $\pi$-circuits per structure as indicated in Fig. \ref{resonances}(a). Here first for anthracene, with 1 or 2 Li atoms each with a low electronegativity, one can imagine that they tend to lose their valence electrons to the anthracene.  And granted electron addition to the anthracene 6-network it should be at the more reactive "radicaloid" sites, which can be anticipated to be the two sites located on a central reflection plane because the 2 end rings then manifest independent local benzene-like resonance, as indicated in Fig. \ref{resonances}, that is, the p-orbitals on these central sites tend toward doubly occupied (from transfer of an electron from the electropositive Li atom or atoms), whence the double bonding to these sites becomes weakened and the two end rings can more readily independently accommodate alternating single and double bonds as with conjugated 6-circuits per structures as in Fig. \ref{resonances}$(b)$.  There are four Kekul\'e structures, each with two conjugated 6-circuits. Resonance in these end rings is thus more effective. Adsorption in either end rings is less effective as then we would only leave three resonant structures rather than four, see Fig. \ref{resonances}$(c)$. This argument is similarly true for single and double adsorption to the sites in the mirror plane, as the charge on the sites will simply be increased. This agrees with the DFT calculations discussed later. 

As a polyacene becomes ever longer, there is an ever greater enhancement of resonances, with a preference for unpairing to appear at a ring at or near the center.  (If for a polyacene of  $n$ hexagons, the 2 unpaired sites are placed in the $m^{th}$ ring ($m\leq n$), then instead of 1 Clar-sextet \cite{clar,debklein} there arise 2, and instead of  $n+1$ neighbor-paired resonance structures, there occur $m(n-m+1)$).  The $\pi$-bond orders to the central charged C atoms should diminish and give longer bonds.  Moreover, with negatively charged C atoms (participating less so in the rest of the $\pi$-system), one might anticipate $sp^3$ hybridization (with an accompanying bond angle distortion, so that the end rings of anthracene are no longer co-planar).  This distortion might also reasonably arise from the attraction between the negatively charged C atoms and the positively charged Li atoms.  For higher polyacenes, the interruption of the $\pi$-network then again facilitates resonance to the two separated pieces best when the added electrons are near the center.

\begin{figure}
        \centering
                      \includegraphics[width=.4\textwidth]{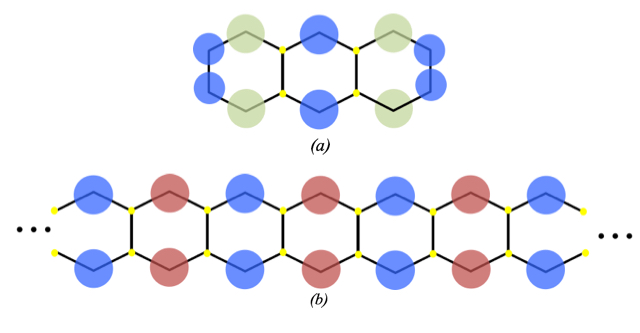}
     \caption{ The LUMO density for anthracene in the first line. And in the second line, the amplitude for a non-bonding MO of infinite polyacene. Different Colors represent different phases.}\label{lumo_infy}
\end{figure}

\subsection{H\"uckel Molecular Orbital Theory Argument}

	The H\"uckel (or simple tight-binding) model for a polyacene of a general number $h$ of hexagons is exactly soluble, as has long been realized \cite{coulnet}.  In this case, the electron(s) from the Li atom(s) should be transferred into the LUMO of neutral anthracene.  Especially for longer polyacenes this LUMO has dominant density on the secondary C atoms (i.e., those at the "points" of the polyacene) concentrated more-so at these C atoms nearer the center - as indicated in Fig. \ref{lumo_infy}$(a)$.  For a very long polyacene, it is in fact readily seen that an orbital indicated in Fig. \ref{lumo_infy}$(b)$ has a 0-energy of orbital for the H\"uckel model, and with the whole eigenspectrum (via the Coulson-Rushbrooke theorem \cite{coulson}) being symmetric about this position, it is seen that such an MO be singly occupied non-bonding. The symmetric and antisymmetric combinations of the 2 such orbitals on the top and bottom edges of a sufficiently long polyacene are the HOMO and LUMO, respectively.  Of course, for a finite polyacene, there is some modification to these MOs, but it is reasonable to assume that the LUMO has a density not too far from this.  Thus, for polyacenes in general, the excess unpaired (or weakly paired) electron density ends up in the same places as for the resonance-theoretic argument. The remaining discussion, including geometric bending in the central area, still applies in somewhat the same way.  The diminishment of the bond order for the more central C atoms (at the outer "points" of the hexagons) occurs because that is where the LUMO (and HOMO) tend to be localized.  Again, Li atoms are favored to lie above or below the anthracene plane to enhance orbital overlap between the Li 2$s$ orbital and anthracene 2$p_z$ orbitals. Thus, qualitative predictions, including the tendency of the 2 central C atoms toward $sp^3$ hybridization, are pretty much the same as for resonance theory.
	
For general zigzag-edged strips the tight-binding approach is known \cite{tbz} to give non-bonding MOs. This is consistent with the resonance theoretic result giving unpaired electrons near such edges.

\subsection{ Ab initio Density Functional Theory (DFT) Computations}

\begin{figure}[htbp]
\centering
\includegraphics[width=0.3\textwidth]{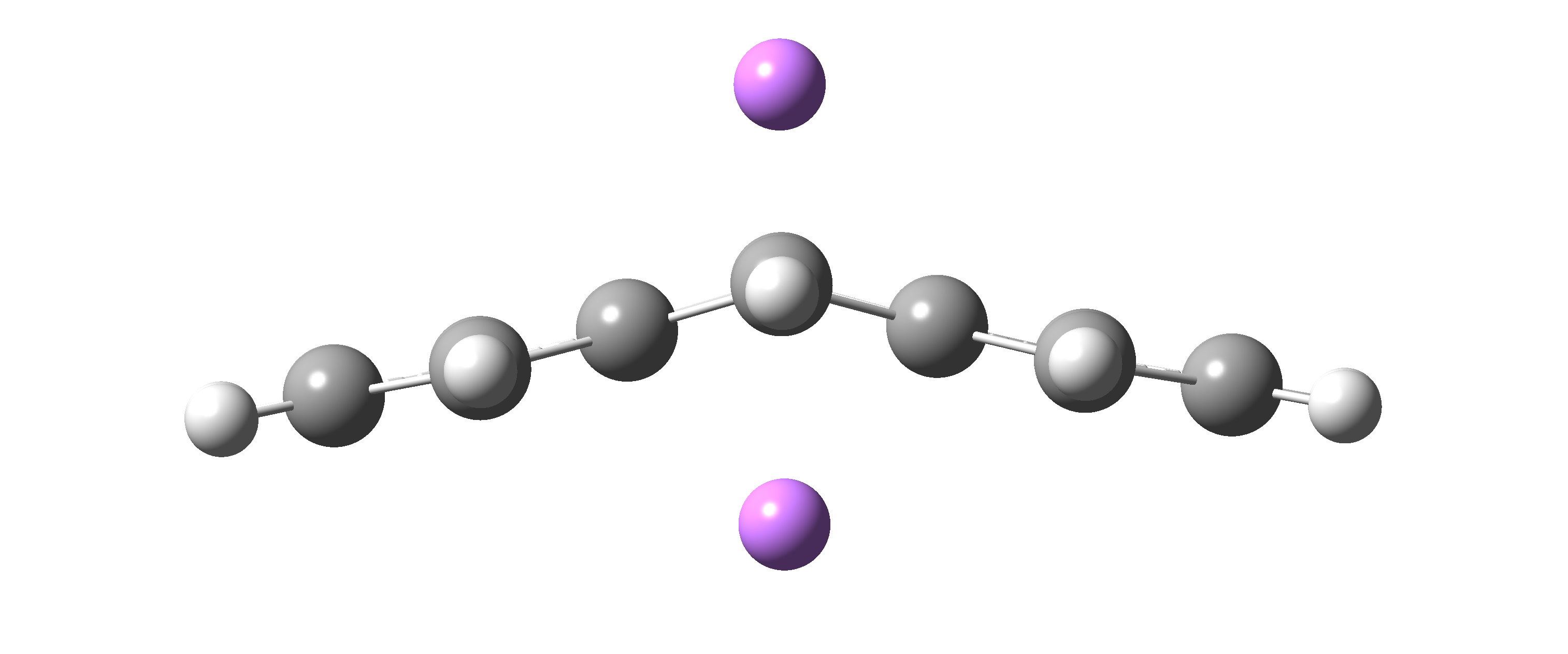}
\includegraphics[width=0.3\textwidth]{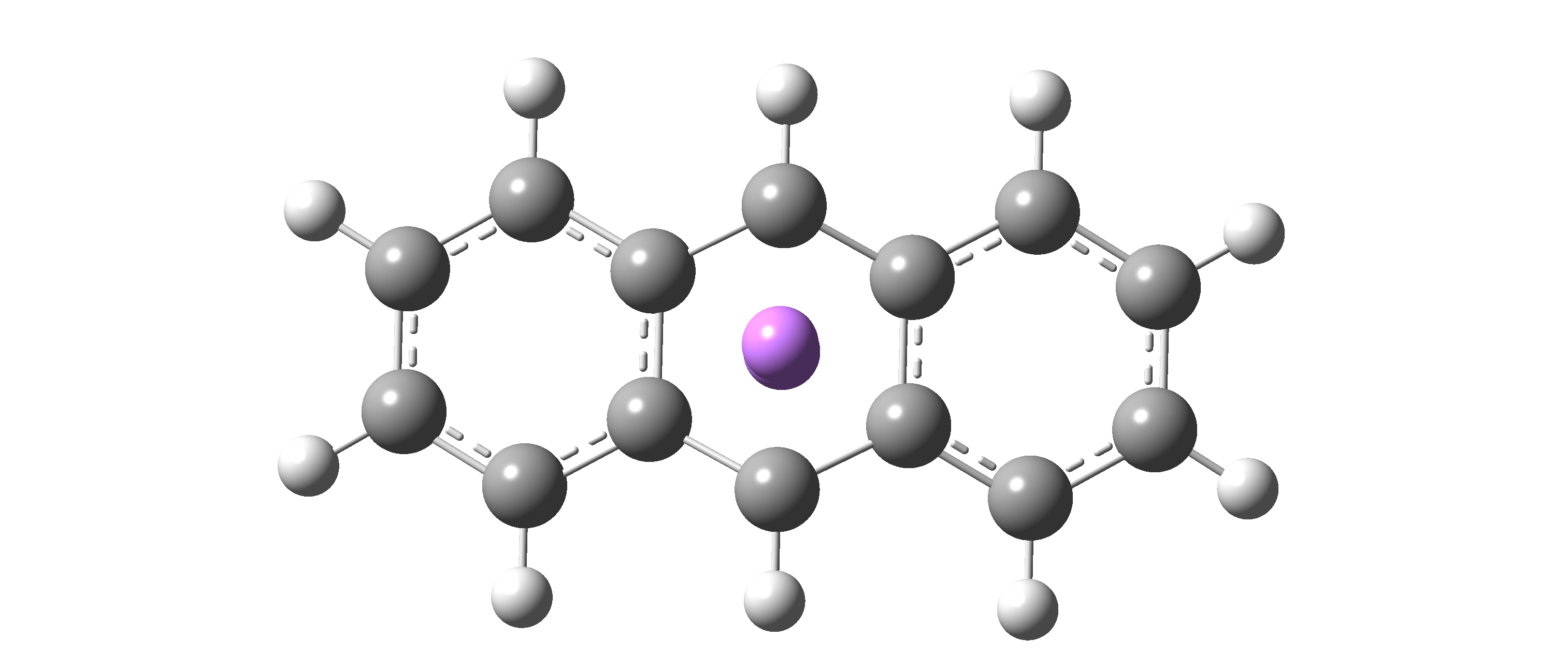}
\caption{Optimized configurations of two Li atoms adsorbed on anthracene obtained using the B3LYP/6-311G* functional and basis set.}
\label{liad}
\end{figure}

In a previous work \cite{ortsel}, we have performed DFT calculations in the software Gaussian 09\cite{gaussian} to obtain the minimum energy configuration of the adsorption of two lithium atoms on opposite sides of anthracene. For those, we used the functional B3LYP\cite{b3lyp1,b3lyp2} and basis set 6-311g adding a polarization function ($d$),  to take into account the effect of the adsorbed alkaline. The resultant geometrical configuration is shown in Fig. \ref{liad}. We have made similar calculations using M\"oller Plesset \cite{moller} perturbation theory (MP2) and basis set 6-311++g** from which we have obtained qualitatively similar results.

\begin{figure}
        \centering
               \includegraphics[width=0.3\textwidth]{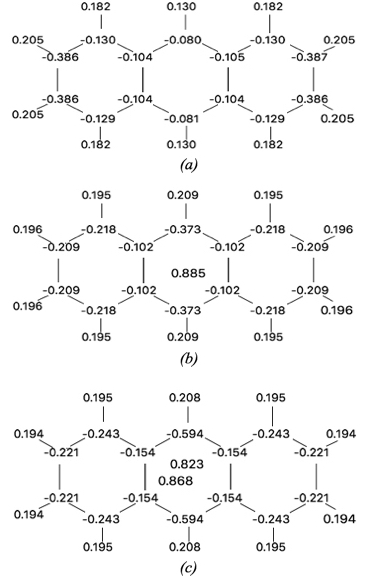}   
     \caption{ Charge distribution of $(a)$ antracene,  $(b)$ one lithium atom adsorbed to anthracene, $(c)$ two lithium atoms adsorbed to anthracene. The lithium atom in the up-center corresponds to the Lithium above the anthracene as in Fig. \ref{liad} and the lithium atom in the down-center corresponds to the lithium below the anthracene.}\label{charges}
\end{figure}

We also performed a charge distribution analysis NAO\cite{nbo} for each case considered in this paper. The excess electron densities near the carbon atoms, as predicted by our resonance and H\"uckels arguments, are consistent with the results of DFT calculations. These calculations indicate that the charge on the lithium atoms is 0.823 for the lithium at the top and 0.868 for the lithium at the bottom, as shown in Fig. \ref{charges}$(c)$.  Thus, there is almost a complete electron transfer (on average) and the simplicities of the qualitative resonance and H\"uckel-MO arguments are plausible.  The bending of the anthracene moiety can be anticipated, both from the resonance- or MO-theoretic- view -points, with excess electron density transferred from the Li atoms more dominantly localized on the two central C atoms, as also shown in Fig. \ref{charges}. These two atoms tend toward (nonplanar) $sp^3$ hybridization, with contributions from a doubly occupied $sp^3$ lone pair. The quantum-mechanical computations clearly give more detailed information, but the general location of the Li atoms and the qualitative aspects of the geometric distortions are borne out, thereby giving some confidence in our simple pictures.

Note that diamonic alkali metal salts of anthracene, tetracene and pentacene are experimentally known \cite{polexp}.

\section{Li Adsorption on Extended Benzenoid Species}

After success with a simple qualitative understanding of Li adsorption on finite polyacenes, it is natural to try to understand how these ideas might apply to extended systems: infinite polyacene zig-zag boundary graphene strips, semi-infinite graphene with a zig-zag boundary, infinite graphene; and multi-layer graphite.

\subsection{Infinite polyacene}

\begin{figure}[htbp]
\centering
\subfloat[Frontal view]{%
  \includegraphics[width=0.45\textwidth]{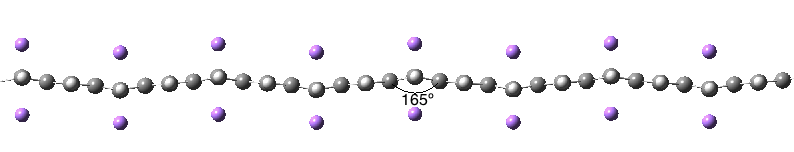}
  \label{fig:periodic_a}
}
\hfill
\subfloat[(Lateral view]{%
  \includegraphics[width=0.45\textwidth]{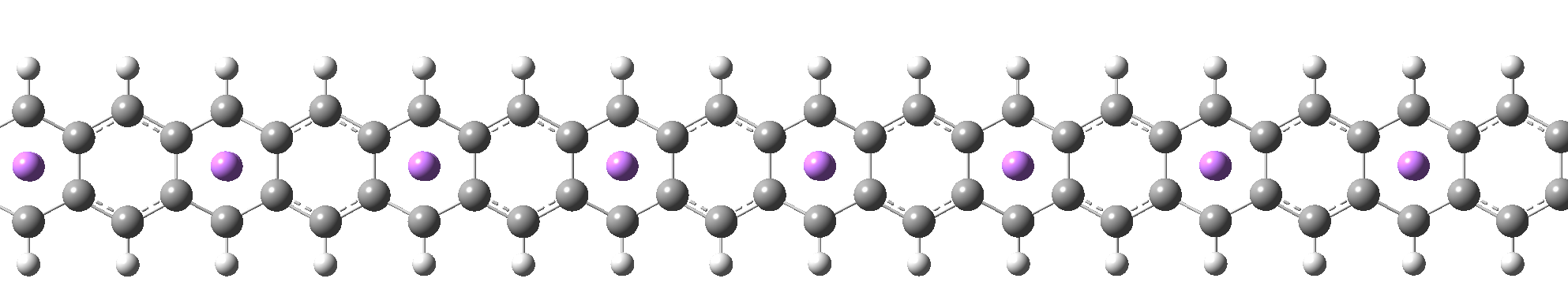}
  \label{fig:periodic_b}
}
\caption{Periodic calculations of the unit cell containing four pairs of Li atoms obtained using the B3LYP functional and the 6-311G basis set.}
\label{periodic}
\end{figure}

If Li atoms are placed periodically along a polycene strip, one obtains a system with a finite unit cell, though to leave some chance for resonance the Li atoms should skip at least one hexagonal ring.  In terms of the MO approach, there are polyacene bands (from above and below the Fermi energy) merging into the Fermi-energy $\varepsilon_F$ ( = 0 for our simple MO model), so that the unoccupied band orbitals nearer this Fermi energy are preferably occupied by any additional electrons. That is, without occupancy of the whole band, one imagines unsubstituted rings between the Li-perturbed ones.  From the resonance-theoretic point of view, the polyradicality increases proportionally to the polyacene length \cite{polyradical}. Our DFT computations \cite{tesis} reveal a favored structure as in Fig. \ref{periodic}, where a Peierls sort of transition occurs \cite{peierls}, in which, because the polyacene is bent, the period of the chain is augmented compared with the flat case. The results there again are in consonance with our qualitative resonance and MO-theoretic arguments.

\subsection{ Semi-infinite Graphene with Zig-Zag Edges}

Both the resonance and H\"uckel theoretical approaches seemingly become somewhat more involved, because of the extended size of the system. Nevertheless, each approach\cite{kleinbyt} can be applied to yield much the same prediction: unpaired (or nonbonding), electron density occurs localized along the zig-zag edge to the extent that there is 1/3 of an unpaired (or singly occupied MO) electron per unit cell of edge. Moreover, while all these unpaired electrons exhibit an exponential density decay away from the edge, the unpaired spin density for different electrons (for nonbonding  MOs) have a varying range of penetrations into the bulk of the graphene and thereby are not mutually back Fourier transformable to longitudinally localized orbitals. As such, these mutually orthogonal band MOs have extensive differential overlap and ferromagnetic coupling amongst these nominally unpaired spins. Observations supporting all of this were made by both resonance-theoretic \cite{seitzRes} and MO-theoretic studies\cite {steinMO, fujitaMO}. More comprehensive tight-binding computations \cite{tb1,tb2,tb3} and some experimental results \cite{exp1, exp2} further support this picture. The resonance-theoretic arguments indeed predict that for a general (translationally symmetric)  edge the number of unpaired electrons per unit cell of edge is $u=|(n_{1\star}+2n_{2\star})-(n_{1\circ}+2n_{2\circ})|/3$ where $n_{d\star}$ and $n_{d\circ}$ are the respective numbers of degree $d$ "starred" or "unstarred" sites per unit cell of edge.  

The overall result then is that there are edge-localized LUMOs to accept electrons donated from nearby Li atoms, again above or below the graphene plane. In addition to the edge-localized nonbonding MOs it is well known \cite{kleinbyt} that bulk graphene has bulk bend orbitals at and near the Fermi level, through a rather low density of them at the apex of a Dirac cone. Thus, Li atoms should stick to the bulk surface of graphene, albeit either less tightly if at the same inter-Li spacings as at the edge, or else at a lesser density (i.e. a larger inter-Li spacing) if at the same strength of binding as  at the edge.

\subsection{ Zig-zag-boundary Graphene Strips}

\begin{figure}[htbp]
\centering
\subfloat[Frontal view]{%
  \includegraphics[height=0.15\textheight]{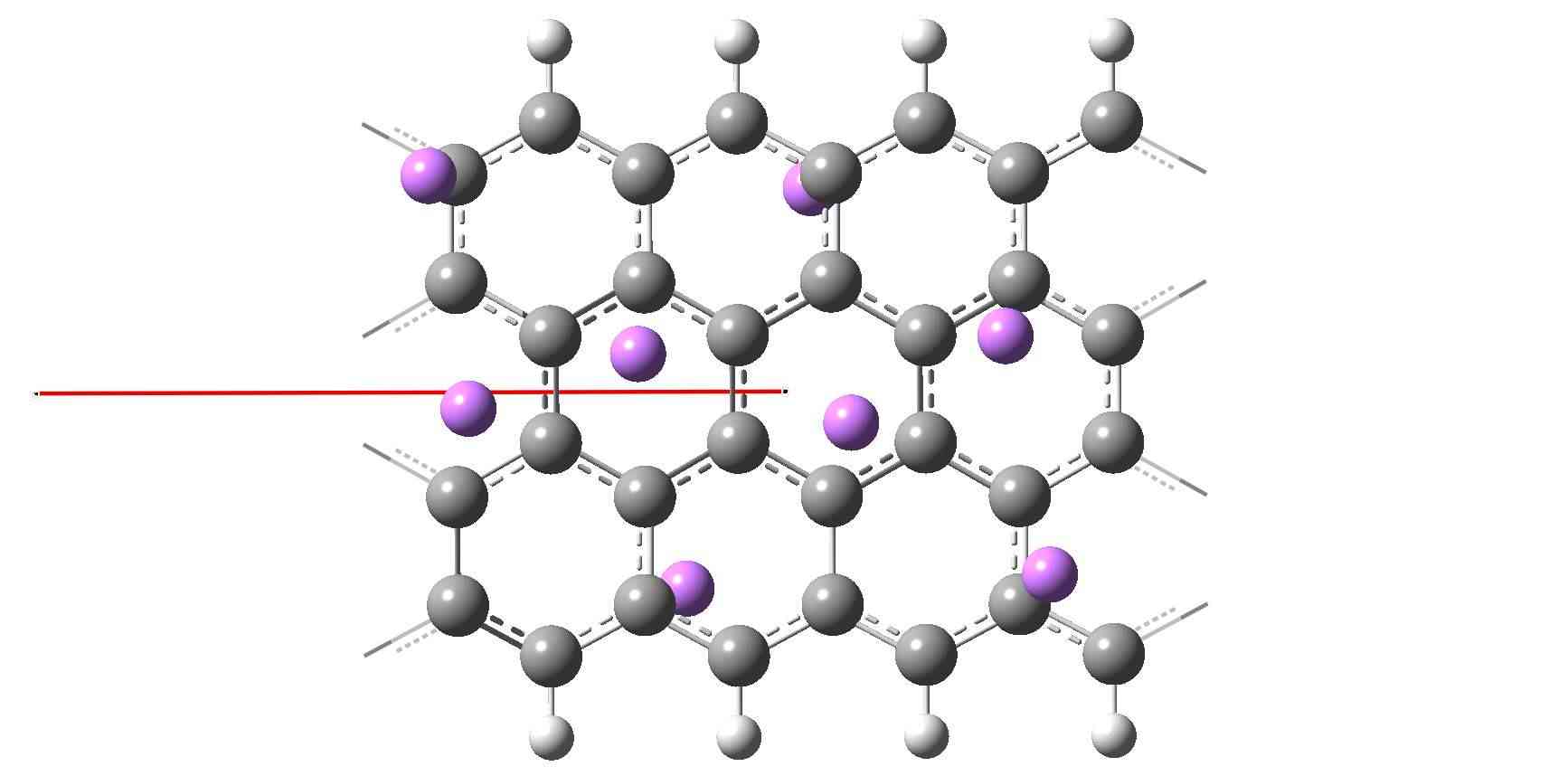}
 \label{fig:period_min_a}
}
\hfill
\subfloat[Lateral view]{%
  \includegraphics[height=0.15\textheight]{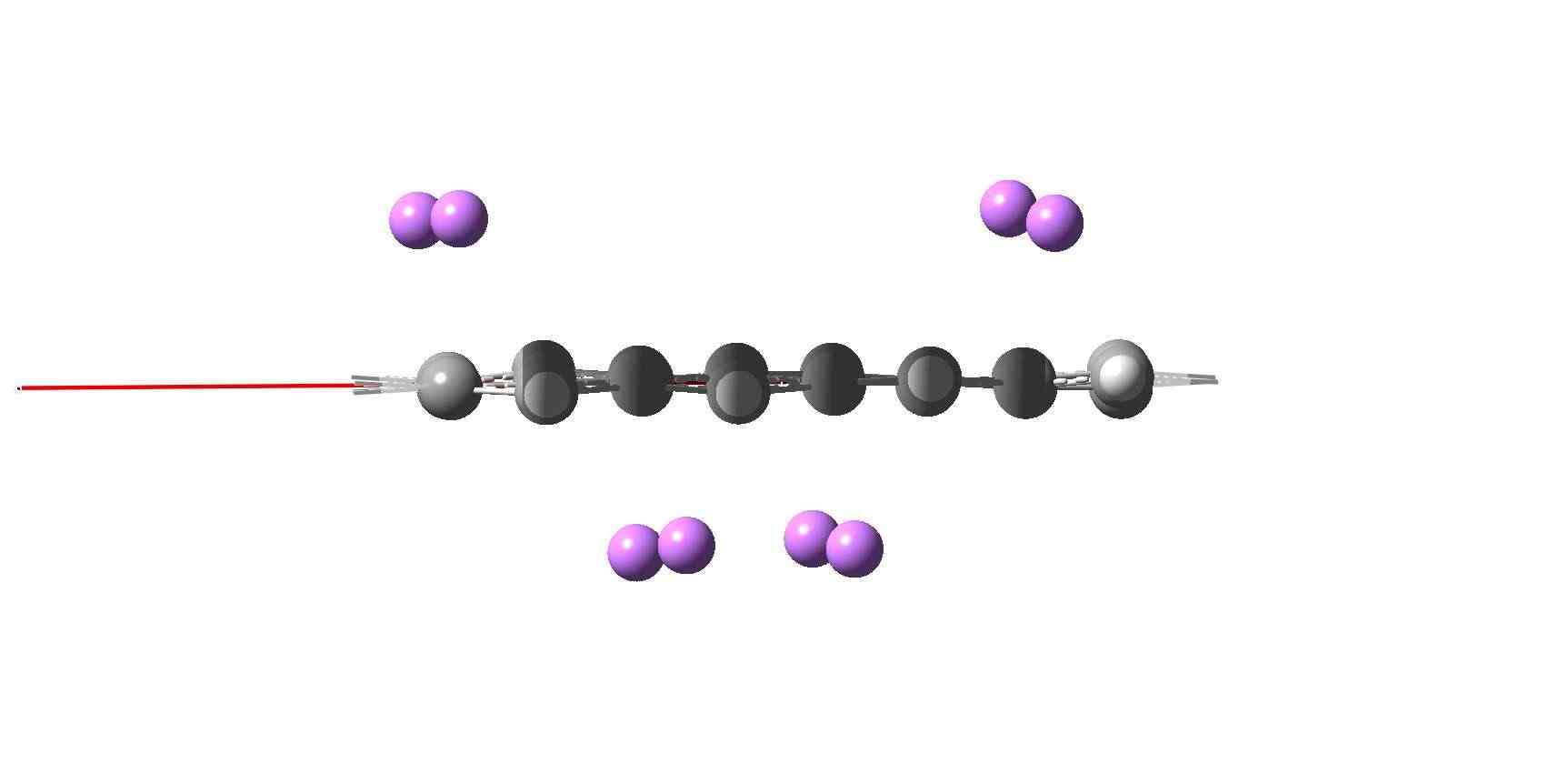}
  \label{fig:period_min_b}
}
\caption{Periodic DFT calculations of the unit cell containing four pairs of Li atoms using the B3LYP functional and the 6-31G basis set.}
\label{period_min}
\end{figure}

Calculations on semi-infinite graphene were done with two parallel zig-zag edges at each of which one anticipates greater unpaired electron density. At higher densities of Li atoms, adsorption to the interior of the strip can be expected. A DFT computation sharing this is found in Fig. \ref{period_min}. A sort of semi-infinite graphene remains essentially flat. It still prefers to occur in pairs above and below the graphene net. The results here again are consonant with our qualitative resonance and MO-theoretical arguments.

\subsection{Expected Results for Larger Systems}

For graphene, following our discussion of semi-infinite graphene, one can expect bulk adsorption. The structure should be close to flat because the rings at which the Li atoms are adsorbed are within linear sequences of polyacene chains in 3 different equivalent directions. To minimize Coulomb repulsion from (partially) ionic Li atoms, one can anticipate the Li atoms to preferably adsorb half on each side of the graphene sheet, end those on the same side are not to near to one another.

In graphite one expects the Li atom between 2 successive sheets to bind  the sheets more strongly than in ordinary  graphite. 

\subsection{Implications for Li-Storage and 2D Carbon Functionalization}

The qualitative arguments developed in the preceding sections-based on resonance theory, H\"uckel molecular-orbital analysis, and density-functional computations-have direct implications for present research on lithium storage in carbon-based nanomaterials. Recent ab initio and experimental studies confirm that the electronic behavior anticipated for Li-decorated polyacenes and zig-zag graphene strips underlies the performance of graphene-derived anodes in Li-ion batteries \cite{aljaber2025}.

As observed here, lithium adsorption preferentially occurs at sites of high $\pi$-electron localization, such as the outer "point" carbons in polyacenes or along zig-zag graphene edges, where the electron density exhibits partial radicaloid character. These regions promote near-complete charge transfer from Li to the carbon $\pi$-network, leading to local weakening of C=C bonds and a partial transition toward sp\textsuperscript{3} hybridization. This effect corresponds to the structural distortions detected in our DFT geometries and is consistent with the local rehybridization and charge redistribution reported in periodic DFT analyses of Li-adsorbed graphene \cite{gong2021}.

In terms of band-structure modification, Li adsorption introduces occupied states near the Fermi level, effectively reducing or closing the intrinsic band gap in extended benzenoid systems. For polyacene-like chains, this electron injection resembles n-type doping and enhances electronic conductivity, a property central to high-rate performance in Li-ion cells. At higher Li concentrations, local pairing of adsorbed Li atoms above and below the carbon plane-predicted both here and by later periodic calculations-stabilizes the lattice while maintaining metallic conductivity.

From the broader perspective of materials functionalization, these findings illustrate how molecular concepts such as Kekul\'e resonance and edge localization remain powerful descriptors of electronic behavior in modern two-dimensional systems. The theoretical framework established for finite polyacenes thus bridges the molecular and mesoscopic limits of Li-decorated graphene, providing a mechanistic rationale for the enhanced binding and charge-storage capacity observed in graphene nanoplatelets and other 2D carbons. In this way, the interplay between resonance stabilization, charge transfer, and edge reactivity unifies classical chemical intuition with the current design principles for nanostructured energy materials.

\section{Conclusions and outlook}

A comprehensive qualitative interpretation of lithium adsorption on anthracene, general polyacenes, and graphene systems with zig-zag boundaries has been presented using three complementary approaches: classical resonance theory, H\"uckel molecular-orbital analysis, and ab initio DFT computations. The convergence of these viewpoints demonstrates that even simple qualitative frameworks can successfully predict the main electronic and structural consequences of alkali-metal adsorption on $\pi$-conjugated carbon systems.

For finite polyacenes, resonance-theoretic and H\"uckel models predict that electron transfer from Li preferentially occurs at central "radicaloid" carbon sites, where bond weakening and local sp\textsuperscript{3} distortion promote independent aromatic stabilization of the terminal rings. The DFT calculations confirm this picture, showing nearly complete charge transfer and geometric bending consistent with localized excess electron density. As the molecular size increases toward the infinite-polyacene or graphene limit, the localization of unpaired $\pi$-electrons evolves into edge-centered nonbonding states, accounting for the enhanced chemical reactivity and charge-accepting capability of zig-zag terminations.

These findings, when viewed in the light of recent theoretical and experimental studies \cite{aljaber2025, gong2021}, reveal that the molecular mechanisms described here extend seamlessly to the nanoscale and even to macroscopic 2D systems. The local rehybridization, charge redistribution, and metallic character predicted for Li-decorated polyacenes are now recognized as key contributors to the high capacity and rate performance of graphene-based anodes in Li-ion batteries. Thus, concepts originating from resonance theory-Kekul\'e structures, Clar sextets, and radicaloid edge sites-remain relevant descriptors for understanding and designing functionalized graphene and other carbon nanomaterials.

In summary, the consistency among classical chemical reasoning, tight-binding theory, and modern DFT not only validates the qualitative predictions made for Li adsorption but also underscores the enduring value of conceptual simplicity in interpreting the complex electronic behavior of emerging nanostructured materials.

Yet there remains an interesting aspect to be explored. As we can see in Fig. 4, a DFT calculation shows that there is a strong spontaneous symmetry breaking occurring with double adsorption of lithium on the central ring of anthracene. Indeed, there have been two previous papers involving two of the authors \cite{jalbout2009,spontansymm2013}, where such a symmetry breaking with double adsorption is shown on polyacene strips and graphene flakes. These studies involved extensive DFT calculations, but the reason for these results has not been discovered. Hopefully, the two arguments, which do not involve infinite chains, will solve the riddle.


\begin{acknowledgements}

The authors acknowledge financial support from the Welch Foundation (Grant BD-0894, Houston, Texas) and from SECIHTI FRONTERAS (Grant CF-2023-G-763). Y.P. Ortiz gratefully acknowledges a postdoctoral fellowship from CONACyT and additional support from research grants 219993 and PAPIIT-DGAPA-UNAM (Grant IG100616). Computational resources were provided by the MIZTLI supercomputing facility under project LANCAD-UNAM-DGTIC-016.

\end{acknowledgements}

\nocite{*}
\bibliographystyle{aipnum4-1}
\bibliography{References}

\end{document}